\documentclass[useAMS,usenatbib,usegraphicx]{mn2e}
\usepackage{amsmath,epsfig,graphicx,color}
\title[X-rays from SN 2012ca]{X-ray Emission from SN 2012ca: A Type Ia-CSM Supernova Explosion in a Dense Surrounding Medium}

\author[Bochenek, C. ~D., Dwarkadas, V. V., Silverman, J. M., et al.]{Christopher D.~Bochenek,$^{1,2}\thanks{E-mail: cbochenek@astro.caltech.edu}$,
V. V. Dwarkadas$^{1}\thanks{E-mail: vikram@oddjob.uchicago.edu}$, 
Jeffrey M.~Silverman$^{3}$, 
Ori D. Fox$^{4}$,
\newauthor
Roger A. Chevalier$^{5}$,
Nathan Smith$^{6}$,
and Alexei V. Filippenko$^{7,8}$
\\
$^{1}$Department of Astronomy and Astrophysics, University of Chicago, 5640 S Ellis Ave, Chicago, IL 60637, USA\\
$^{2}$Current address: Astronomy Department, California Institute of Technology, 1200 E. California Boulevard, Pasadena, CA 91125, USA\\
$^{3}$Department of Astronomy, University of Texas, Austin, TX 78712, USA\\
$^{4}$Space telescope Science Institute, Baltimore, MD 21218, USA\\
$^{5}$Department of Astronomy,University of Virginia, Charlottesville, VA 22903, USA\\
$^{6}$Steward Observatory, 933 N. Cherry Ave., Tucson, AZ 85721, USA \\
$^{7}$Department of Astronomy, University of California, Berkeley, CA 94720-3411, USA \\
$^{8}$Senior Miller Fellow, Miller Institute for Basic Research in Science,
University of California, Berkeley, CA 94720, USA \\
}

\begin{document}
\newcommand{\vper}{\mbox{${v_{\perp}}$}}
\newcommand{\vpar}{\mbox{${v_{\parallel}}$}}
\newcommand{\uper}{\mbox{${u_{\perp}}$}}
\newcommand{\vperout}{\mbox{${{v_{\perp}}_{o}}$}}
\newcommand{\uperout}{\mbox{${{u_{\perp}}_{o}}$}}
\newcommand{\vperin}{\mbox{${{v_{\perp}}_{i}}$}}
\newcommand{\uperin}{\mbox{${{u_{\perp}}_{i}}$}}
\newcommand{\upar}{\mbox{${u_{\parallel}}$}}
\newcommand{\uparout}{\mbox{${{u_{\parallel}}_{o}}$}}
\newcommand{\vparout}{\mbox{${{v_{\parallel}}_{o}}$}}
\newcommand{\uparin}{\mbox{${{u_{\parallel}}_{i}}$}}
\newcommand{\vparin}{\mbox{${{v_{\parallel}}_{i}}$}}
\newcommand{\dout}{\mbox{${\rho}_{o}$}}
\newcommand{\din}{\mbox{${\rho}_{i}$}}
\newcommand{\da}{\mbox{${\rho}_{1}$}}
\newcommand{\mfast}{\mbox{$\dot{M}_{f}$}}
\newcommand{\mslow}{\mbox{$\dot{M}_{a}$}}
\newcommand{\beqn}{\begin{eqnarray}}
\newcommand{\eeqn}{\end{eqnarray}}
\newcommand{\be}{\begin{equation}}
\newcommand{\ee}{\end{equation}}
\newcommand{\noi}{\noindent}
\newcommand{\ftheta}{\mbox{$f(\theta)$}}
\newcommand{\gtheta}{\mbox{$g(\theta)$}}
\newcommand{\ltheta}{\mbox{$L(\theta)$}}
\newcommand{\stheta}{\mbox{$S(\theta)$}}
\newcommand{\utheta}{\mbox{$U(\theta)$}}
\newcommand{\xitheta}{\mbox{$\xi(\theta)$}}
\newcommand{\vs}{\mbox{${v_{s}}$}}
\newcommand{\ro}{\mbox{${R_{0}}$}}
\newcommand{\pa}{\mbox{${P_{1}}$}}
\newcommand{\va}{\mbox{${v_{a}}$}}
\newcommand{\vo}{\mbox{${v_{o}}$}}
\newcommand{\vp}{\mbox{${v_{p}}$}}
\newcommand{\vw}{\mbox{${v_{w}}$}}
\newcommand{\vf}{\mbox{${v_{f}}$}}
\newcommand{\lprime}{\mbox{${L^{\prime}}$}}
\newcommand{\uprime}{\mbox{${U^{\prime}}$}}
\newcommand{\sprime}{\mbox{${S^{\prime}}$}}
\newcommand{\xiprime}{\mbox{${{\xi}^{\prime}}$}}
\newcommand{\mdot}{\mbox{$\dot{M}$}}
\newcommand{\msun}{\mbox{$M_{\odot}$}}
\newcommand{\yr}{\mbox{${\rm yr}^{-1}$}}
\newcommand{\kms}{\mbox{${\rm km} \;{\rm s}^{-1}$}}
\newcommand{\lambdav}{\mbox{${\lambda}_{v}$}}
\newcommand{\lequ}{\mbox{${L_{eq}}$}}
\newcommand{\eqpratio}{\mbox{${R_{eq}/R_{p}}$}}
\newcommand{\ra}{\mbox{${r_{o}}$}}
\newcommand{\bfig}{\begin{figure}[h]}
\newcommand{\efig}{\end{figure}}
\newcommand{\tone}{\mbox{${t_{1}}$}}
\newcommand{\done}{\mbox{${{\rho}_{1}}$}}
\newcommand{\dsn}{\mbox{${\rho}_{SN}$}}
\newcommand{\dzero}{\mbox{${\rho}_{0}$}}
\newcommand{\ve}{\mbox{${v}_{e}$}}
\newcommand{\vej}{\mbox{${v}_{ej}$}}
\newcommand{\Mch}{\mbox{${M}_{ch}$}}
\newcommand{\mej}{\mbox{${M}_{e}$}}
\newcommand{\Mst}{\mbox{${M}_{ST}$}}
\newcommand{\dam}{\mbox{${\rho}_{am}$}}
\newcommand{\Rst}{\mbox{${R}_{ST}$}}
\newcommand{\Vst}{\mbox{${V}_{ST}$}}
\newcommand{\Tst}{\mbox{${T}_{ST}$}}
\newcommand{\no}{\mbox{${n}_{0}$}}
\newcommand{\Efif}{\mbox{${E}_{51}$}}
\newcommand{\rsh}{\mbox{${R}_{sh}$}}
\newcommand{\msh}{\mbox{${M}_{sh}$}}
\newcommand{\vsh}{\mbox{${V}_{sh}$}}
\newcommand{\vrev}{\mbox{${v}_{rev}$}}
\newcommand{\rpr}{\mbox{${R}^{\prime}$}}
\newcommand{\mpr}{\mbox{${M}^{\prime}$}}
\newcommand{\vpr}{\mbox{${V}^{\prime}$}}
\newcommand{\tpr}{\mbox{${t}^{\prime}$}}
\newcommand{\cone}{\mbox{${c}_{1}$}}
\newcommand{\ctwo}{\mbox{${c}_{2}$}}
\newcommand{\cthree}{\mbox{${c}_{3}$}}
\newcommand{\cfour}{\mbox{${c}_{4}$}}
\newcommand{\Te}{\mbox{${T}_{e}$}}
\newcommand{\Ti}{\mbox{${T}_{i}$}}
\newcommand{\Ha}{\mbox{${H}_{\alpha}$}}
\newcommand{\Rprime}{\mbox{${R}^{\prime}$}}
\newcommand{\Vprime}{\mbox{${V}^{\prime}$}}
\newcommand{\Tprime}{\mbox{${T}^{\prime}$}}
\newcommand{\Mprime}{\mbox{${M}^{\prime}$}}
\newcommand{\rprime}{\mbox{${r}^{\prime}$}}
\newcommand{\rfprime}{\mbox{${r}_f^{\prime}$}}
\newcommand{\vprime}{\mbox{${v}^{\prime}$}}
\newcommand{\tprime}{\mbox{${t}^{\prime}$}}
\newcommand{\mprime}{\mbox{${m}^{\prime}$}}
\newcommand{\Me}{\mbox{${M}_{e}$}}
\newcommand{\nh}{\mbox{${n}_{H}$}}
\newcommand{\rr}{\mbox{${R}_{2}$}}
\newcommand{\rf}{\mbox{${R}_{1}$}}
\newcommand{\vtwo}{\mbox{${V}_{2}$}}
\newcommand{\vout}{\mbox{${V}_{1}$}}
\newcommand{\dshell}{\mbox{${{\rho}_{sh}}$}}
\newcommand{\dwind}{\mbox{${{\rho}_{w}}$}}
\newcommand{\dslow}{\mbox{${{\rho}_{s}}$}}
\newcommand{\dfast}{\mbox{${{\rho}_{f}}$}}
\newcommand{\vfast}{\mbox{${v}_{f}$}}
\newcommand{\vslow}{\mbox{${v}_{s}$}}
\newcommand{\cc}{\mbox{${\rm cm}^{-3}$}}
\newcommand{\apj}{\mbox{ApJ}}
\newcommand{\apjl}{\mbox{ApJL}}
\newcommand{\apjs}{\mbox{ApJS}}
\newcommand{\aj}{\mbox{AJ}}
\newcommand{\araa}{\mbox{ARAA}}
\newcommand{\nat}{\mbox{Nature}}
\newcommand{\aap}{\mbox{AA}}
\newcommand{\gca}{\mbox{GeCoA}}
\newcommand{\pasp}{\mbox{PASP}}
\newcommand{\mnras}{\mbox{MNRAS}}
\newcommand{\apss}{\mbox{ApSS}}

\date{}

\pagerange{\pageref{firstpage}--\pageref{lastpage}} \pubyear{2016}

\maketitle

\label{firstpage}

\begin{abstract}
  {X-ray emission is one of the signposts of circumstellar interaction
    in supernovae (SNe), but until now, it has been observed only in
    core-collapse SNe.  The level of thermal X-ray emission is a
    direct measure of the density of the circumstellar medium (CSM),
    and the absence of X-ray emission from Type Ia SNe has been
    interpreted as a sign of a very low density CSM. In this paper, 
    we report late-time (500--800 days after discovery) X-ray 
    detections of SN 2012ca in {\it Chandra} data. The presence of
    hydrogen in the initial spectrum led to a classification of Type
    Ia-CSM, ostensibly making it the first SN~Ia detected with
    X-rays. Our analysis of the X-ray data favors an asymmetric
    medium, with a high-density component which supplies the X-ray
    emission. The data suggest a number density $> 10^8$ cm$^{-3}$ in
    the higher-density medium, which is consistent with the large
    observed Balmer decrement if it arises from collisional
    excitation. This is high compared to most core-collapse SNe, but
    it may be consistent with densities suggested for some Type IIn or
    superluminous SNe.  If SN 2012ca is a thermonuclear SN, the large
    CSM density could imply clumps in the wind, or a dense torus or
    disk, consistent with the single-degenerate channel. A remote
    possibility for a core-degenerate channel involves a white dwarf
    merging with the degenerate core of an asymptotic giant branch 
    star shortly before the explosion, leading to a common envelope 
    around the SN. }
\end{abstract}


\section{Introduction}
Thermal X-ray emission is one of the clearest indications of
circumstellar interaction in supernovae \citep[SNe;][]{chevalier82,
  cf94}. The intensity of the emission depends on the square of the
density, and thus is a good estimator of the density of the ambient
medium, as long as the emission is not absorbed by the medium. So far,
over 60 SNe have been detected in X-rays \citep{vvd12, vvd14}. All of
them are core-collapse SNe, where the circumstellar medium (CSM) is
formed by mass loss from the progenitor star. Until now, no Type Ia SN
has been detected in X-rays.  Deep limits on the emission from SNe~Ia
in the radio \citep{chomiuketal16} and X-ray \citep{marguttietal14}
bands indicate a very low mass-loss rate ($\le 10^{-9}$ M$_{\odot}$
yr$^{-1}$) for the stellar progenitor system, suggesting that SNe~Ia
are surrounded by a very low density CSM, if they even have one.

It is generally accepted that the progenitor of a SN~Ia must be a
white dwarf. In order to raise the mass of this white dwarf to nearly
the Chandrasekhar limit to produce an explosion, it must have accreted
mass transferred from a companion in a binary system.  The nature of
the companion is hotly debated \citep[e.g.,][]{ruizlapuente14}, with
evidence existing for both double-degenerate and single-degenerate
systems \citep{Maozetal2014}. In the former case, the companion is
another white dwarf
\citep{Scalzoetal2010,Silvermanetal2011,Nugentetal2011,Bloometal2012,Brownetal2012},
while in the latter case it is a main sequence or evolved star
\citep{hamuyetal2003,Dengetal2004}. It is likely that both channels
are present.

Some SNe~Ia exhibit narrow hydrogen lines superimposed on a SN~Ia-like
spectrum \citep{hamuyetal2003,Dengetal2004}. The narrow line width
suggests velocities much lower than that of the expanding shock wave,
and are therefore presumed to arise in the surrounding medium.  This
indicates the presence of an ambient medium, perhaps arising from mass
loss from the companion star.  These SNe comprise the subclass of Type
Ia-CSM. \citet{silvermanetal2013a} composed a list of several common
features of SNe~Ia-CSM.  The absolute magnitudes of SNe~Ia-CSM are
larger than those of normal SNe~Ia. They even exceed those of most
SNe~IIn, whose spectra show relatively narrow lines (hence the ``n''
designation; see \citet{filippenko97} for a review) on a broad base.
The spectra of SNe~Ia-CSM consist of a SN~Ia spectrum diluted by
relatively narrow hydrogen lines, a blue continuum created from many
overlapping broad lines from iron-group elements, and strong H$\alpha$
emission with width $\sim 2000$ km~s$^{-1}$. The H$\alpha$ profile
varies for $\sim 100$ days post-explosion before steadily
increasing. He~I and H$\beta$ emission are observed, but are
relatively weak. { SNe~Ia-CSM have larger Balmer decrements (the
  ratio of the intensity of the H$\alpha$ to H$\beta$ lines) than the
  typical value under interstellar conditions of $\sim2.86$}, most
likely resulting from emission due to collisional excitation rather
than recombination, suggesting high-density CSM shells that are
collisionally excited when the faster-moving SN ejecta overtake
them. SNe~Ia-CSM have never been detected at radio wavelengths, and this
work represents the first detection of a SN~Ia-CSM in X-rays. The host
galaxies of SNe~Ia-CSM appear to be spiral galaxies having
Milky-Way-like luminosities with solar metallicities, or irregular
dwarf galaxies similar to the Magellenic Clouds with subsolar
metallicities.

Since SNe~Ia-CSM exhibit characteristic features of SNe~Ia spectra with H lines
that are the defining aspect of Type II SNe, there is still debate on
their exact nature. Some suggest that they are odd core-collapse SNe
masquerading as a SN~Ia \citep{Benettietal2006}. However, there are
indications that some SNe~Ia-CSM are likely thermonuclear explosions
interacting with a CSM. PTF11kx is by most accounts a SN~Ia
produced from the single-degenerate channel
\citep{Dildayetal2012, silvermanetal2013b}. Observations of PTF11kx show
the evolution of its spectrum to be similar to that of SN 1999aa, a SN~Ia. 
It appears to have a multi-component CSM, with no signs of a CSM 
in the early-time observations. There exists faster-moving material
closer to the SN, and shells of CSM
\citep{Dildayetal2012} expanding outward in radius from the SN. These
features are interpreted by the authors as being consistent with
recurrent nova eruptions, and thus suggest a thermonuclear explosion
of a white dwarf through the single-degenerate channel.

SN 2012ca reignited the debate about whether all SNe~Ia-CSM
are thermonuclear explosions. While a thermonuclear
progenitor is debated for SN 2012ca, its spectral classification is
Type Ia with superimposed relatively narrow H
lines. \citet{Inserraetal2014} argue that SN 2012ca is a
core-collapse SN based on the identification of several
intermediate-mass elements such as C, Mg, and O in its nebular spectrum.
{ Furthermore, \citet{Inserraetal2016} point out that SN 2012ca is
likely a core-collapse SN on the basis of energetics,
suggesting that the conversion of kinetic energy into luminosity must
be between 20\% and 70\%, an unusually large
value. \citet{Foxetal2015} argue that the C, Mg, and O features seen 
by \citet{Inserraetal2014} are misidentified; instead, they suggest that 
those of Mg and O are actually iron lines, and that of C is the Ca~II 
near-infrared (IR) triplet. Based on the lack of broad C, O, and Mg in 
the spectrum of SN 2012ca and the presence of broad iron lines,
\citet{Foxetal2015} conclude that SN 2012ca is more consistent with a
thermonuclear rather than a core-collapse explosion. They further point 
out that the high efficiency required for SN 2012ca is within the realm 
of possibility.}

The goal of this paper is to further study the nature of SN 2012ca via
its X-ray emission. Through the analysis of the X-ray data, we shed
light on the density of the ambient medium and therefore the nature of
the progenitor system. This paper is structured as follows. In \S 2,
we summarize the X-ray data and analysis. In \S 3, we use the optical data
to estimate the SN kinematics. In \S 4, we elaborate on the reduction
and analysis of the spectra. An estimate of the density of the CSM, 
for both a homogeneous and a clumpy medium, is
presented in \S 5. Finally, \S 6 summarizes our work and discusses the
implications of the discovery.

\section{Observations}

SN 2012ca was discovered in the late-type spiral galaxy ESO 336-G009
on 2012 April 25.6 \citep{dpb12}.  The NASA/IPAC Extragalactic
Database (NED) gives the luminosity distance to the galaxy as 80.1 Mpc
for a value of H$_0=73$ km s$^{-1}$ Mpc$^{-1}$, $\Omega_m=0.27$, and
$\Omega_\Lambda=0.73$. We observed SN 2012ca with the \textit{Chandra}
Advanced CCD Imaging Spectrometer (ACIS); see Table
\ref{table:12cadata} for details.  Two observations of 20~ks each were
made about 6 months apart: ObsID 15632 (2013 September 17, hereafter
epoch 1) and 15633 (2014 March 27, hereafter epoch 2). Using an
explosion date of MJD $55998.2 { \pm 20}$ \citep{Inserraetal2016},
these epochs occur 554 days and 745 days after the explosion,
respectively. { All epochs in this paper will be referenced from this
  explosion date,} and all dates are listed in Universal Time (UT).

\begin{table*}
 \begin{minipage}{140mm}
  \caption{Summary of X-ray data on SN 2012ca, listing the satellite
    and instrument which took the observation, the date, the age (days after
    explosion), the exposure time, the count rate, the column density,
    the derived temperature, and the unabsorbed flux, with 1$\sigma$
    error bars where available. }
  \begin{tabular}{@{}llcrrrrr@{}}
  \hline
   Instrument &  Obs Date  & Days After & Exposure  & Count Rate  & $N_{\rm H}$ & $kT$ & 0.5--7~keV \\
 & & Outburst & (ks) & ($10^{-3}$  & ($10^{22}$ & (keV) & Flux ($10^{-14}$ \\
& &  & & counts s$^{-1}$) & ${\rm cm^{-2}}$) & & erg s$^{-1}$ cm$^{-2}$) \\
 \hline
 {\it Chandra} ACIS  & 2013-09-17 & 554 & 20.0 & 1.4 $\pm 0.29$ & 6.0 $\pm 3.5$ &  $2.6 \pm 1.8$ & $7.6_{-5.1}^{+389}$ \\
 {\it Chandra} ACIS  & 2014-03-27 & 745 & 20.0 & $0.29^{+0.26}_{-0.16}$ & $5.2 \pm 2.5$ & $1.2 \pm 0.5$ & $3.7^{+104.0}_{-3.0}$ \\
 \hline
\label{table:12cadata}
\end{tabular}
\end{minipage}
\end{table*}

\begin{figure*}
\includegraphics[scale=0.4]{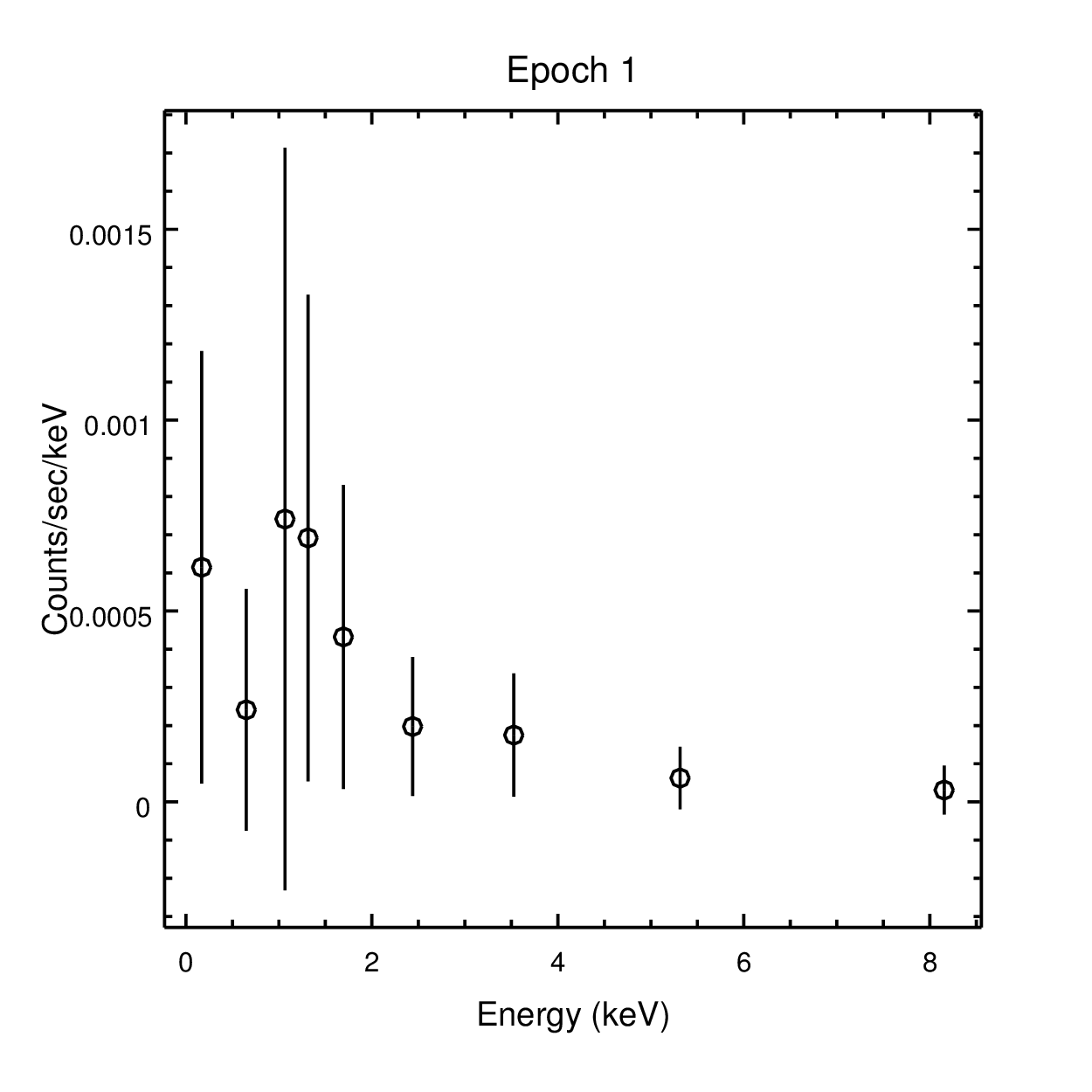}
\includegraphics[scale=0.4]{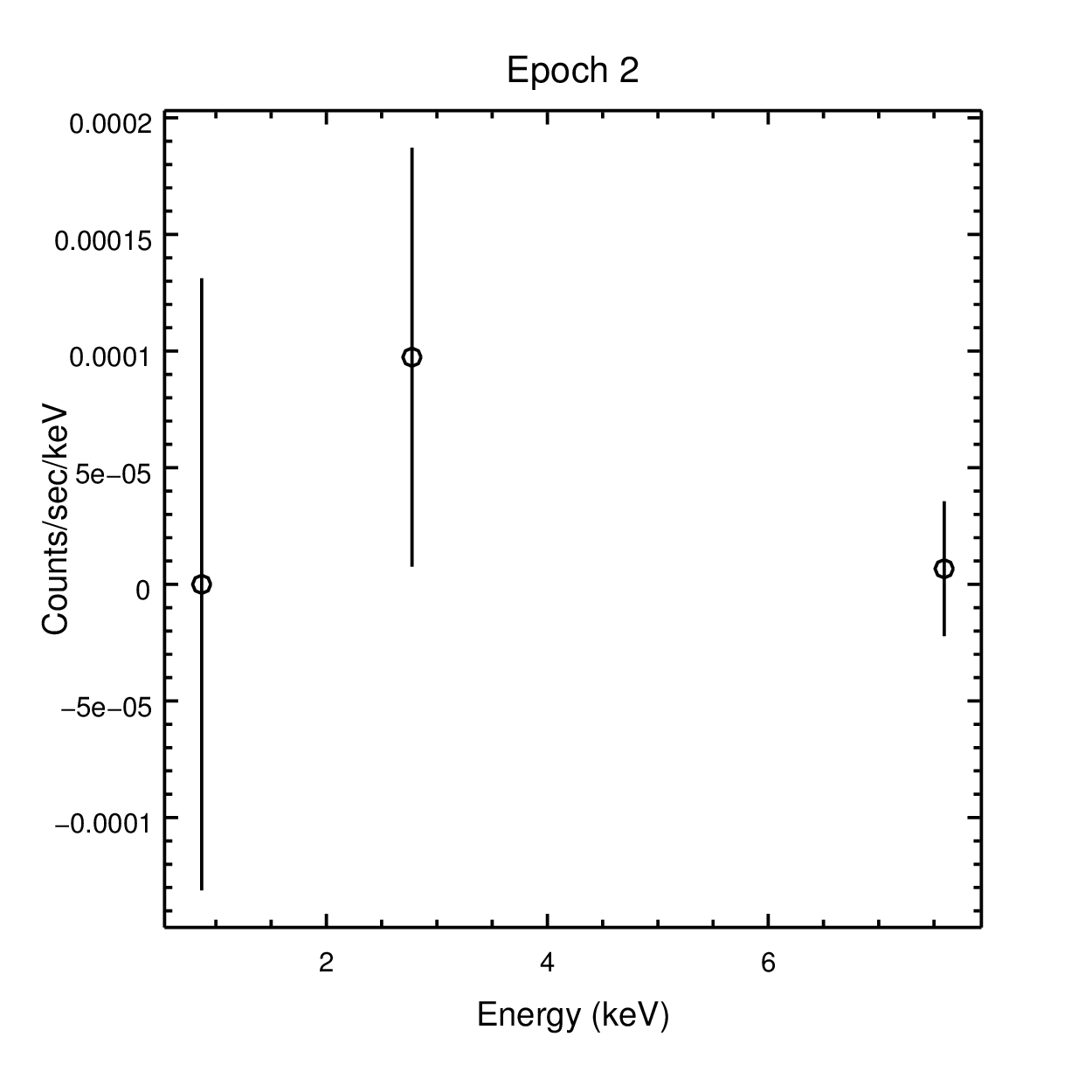}
\caption{Left: X-ray spectrum at epoch 1. The data are grouped such
  that there are 4 counts per bin. Right: Spectrum for epoch 2, also
  grouped into 4 counts per bin. Although the grouped spectra are shown for
  clarity, ungrouped versions of these spectra were used in the
  analysis.}
\label{fig:spectra}
\end{figure*}

The data were analysed using \textit{Chandra} Interactive Analysis of
Observations (CIAO) version 4.7 and CalDB 4.8.  The source and
background regions were each taken to be $4''$ radius, which contains
90\% of the \textit{Chandra} point-spread function. Analysis and
fitting were done using CIAO and Sherpa. Because the number of counts
in each observation is low compared to the background, fitting methods
involving minimizing the $\chi^2$ statistic are invalid. Instead, we
used the ``cstat'' statistic, corresponding to the maximum likelihood
function of the Poisson distribution, and fitted both the data and
background together \citep{Cash1979}. { While we used ungrouped
  spectra in our analysis, the grouped spectra for epochs 1 and 2 are
  shown in Figure \ref{fig:spectra}.}
  
{ We use the two optical spectra of SN 2012ca published by
  \citet{Foxetal2015} at 486 and 508 days after the explosion. The
  most recent spectra, on days 522.8, 548.9, 580.8, were obtained from
  WISeREP \citep{Yaronetal2012} as part of the PESSTO SSDR2
  \citep{Smarttetal2015}. We thank Dr.~Cosimo Inserra for providing
  the rest of the optical spectra of SN 2012ca published by
  \citet{Inserraetal2014}.}

\section{Supernova Kinematics}
\begin{figure*}
\includegraphics[scale=0.3, angle=90]{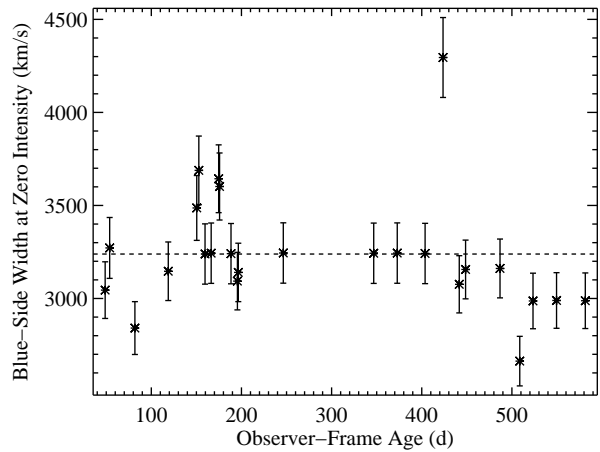}
\caption{Blue-side width at zero intensity (BSWZI) of the H$\alpha$
  line vs.~time since explosion.  The mean velocity indicated is $\sim 3200$ km~s$^{-1}$.}
\label{fig:bswzi}
\end{figure*}

The velocity of the fast-moving gas is required for subsequent
analysis. To determine this, we fit a cubic spline function to each
H$\alpha$ profile in the optical spectra. This spectral fitting
algorithm is similar to the one used by \citet{silvermanetal12}. We
specifically recorded the velocity of the peak of the H$\alpha$
profile, as well as the blue-side width at zero intensity (BSWZI). All
of the H$\alpha$ peaks have measured velocities consistent with 0 km
s$^{-1}$ (in the rest frame of the host galaxy), and the velocity of
the peak does not appear to evolve with time. As shown in Figure
\ref{fig:bswzi}, the BSWZI is also measured to be approximately
constant with time and has a value of $3200 \pm 300$ km s$^{-1}$.

It is not clear if the BSWZI is representative of the shock velocity,
since other factors could contribute to the broadening of the
line. Electron scattering is one such factor, and has been shown to be
relevant in SNe such as SN 2010jl \citep{franssonetal14}. However, it
is unlikely that electron scattering could be important as late as 530
days, when the last optical spectra were obtained. The BSWZI also does
not appear to change substantially in the first 500 days. Furthermore,
the column density that we infer from the X-ray observations, although
large, is still two orders of magnitude lower than that inferred for
SN 2010jl \citep{franssonetal14, chandraetal15}, and would not lead to
large electron scattering depths. Finally, electron scattering results
in Lorentzian line shapes, whereas the line shapes here are better
fitted by a Gaussian. Thus, we do not believe that electron scattering
plays a large role. Other factors such as line blending or
contamination from the reverse shock could be present. Therefore,
although there is evidence of gas moving at velocities as high as 3200
km s$^{-1}$, we cannot easily assume that this represents the shock
velocity.

The Gaussian shape of the lines indicates that there is material in
the SN moving at all velocities from +3500 km s$^{-1}$ to $-3500$ km
s$^{-1}$. We have also measured the full width at half-maximum
intensity (FWHM) of the line from the two high-resolution spectra that
we have, on days 436.2 and 458.1. The FWHM of the H$\alpha$ line is
around 1300 km s$^{-1}$. The other spectra have low resolution, and
measuring the FWHM is quite unreliable, even when corrected for
instrument resolution. It is possible that the lower velocity is more
indicative of the velocity of the bulk of the gas. We address this
issue again in \S 5.

\section{Modeling the Spectrum}
\label{sec:spectrum}

There are several indications of a high density in the surrounding
medium: (1) the large Balmer decrement, which ranges from 3 to 20 for
SN 2012ca, and can potentially be explained by collisional excitation
rather than recombination, as for other Type Ia-CSM; (2) the slow
velocity of the SN shock wave even at early times ($\la 3200$ km
s$^{-1}$), compared to general SN~Ia velocities in excess of $10^4$ km
s$^{-1}$ \citep{wangetal09}; and (3) the high observed X-ray
luminosity (see below) at a deduced redshift of $z=0.019$
\citep{Inserraetal2014}, generally seen in Type IIn SNe with high
densities. In a high-density medium, thermal emission is expected to
dominate. The low statistics make it difficult to robustly distinguish
between nonthermal and thermal models, but the high density suggests
that thermal models are a good starting point. We fit the X-ray
spectrum with the ``vmekal'' model \citep{liedahletal95} using
$z=0.019$, multiplied by the Tuebingen-Boulder model for X-ray
absorption \citep{Wilmsetal2000}. The background was fit with an
8-degree polynomial, and then the background plus data were
simultaneously fit together, using the C-statistic to gauge the
goodness of the fit.

\begin{figure}
\center
\includegraphics[width=\columnwidth]{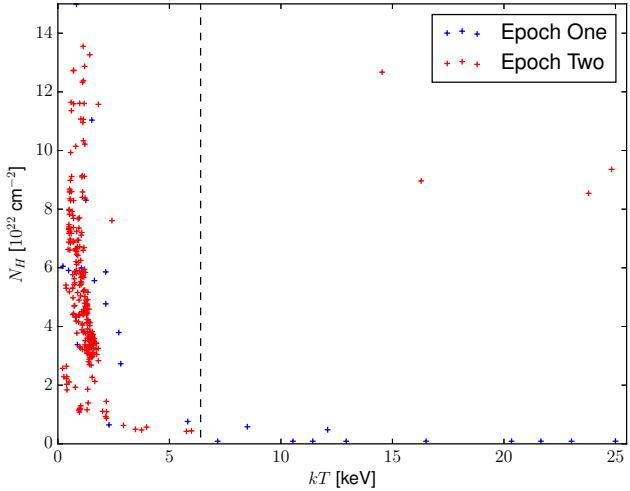}
\caption{The red points (epoch 2) and blue points (epoch 1) represent
  acceptable X-ray spectral fit values for the column density ($N_{\rm
    H}$) and temperature ($kT$) for SN 2012ca, using the method
  outlined in the text. The points to the right of the dashed line are
  excluded from parameter estimation for reasons outlined in the text.
  \label{fig:fits}}
\end{figure}

Given the low statistics, it is difficult to accurately fit a model to
the X-ray spectrum and extract relevant parameters, especially for the
second epoch. Several models with different column densities and
temperatures appear to provide a good fit. For example, we found two
fits to epoch 2 which have reduced statistics of 0.5073 and
0.5071. The former has an $N_{\rm H}$ of $2.65 \times 10^{22}$
cm$^{-2}$ and a $kT$ of 0.38 keV, while the latter has an $N_{\rm H}$
of $4.24 \times 10^{22}$ cm$^{-2}$ and a $kT$ of 1.31 keV. To address
this problem, we explore the likelihood parameter space with a
modified Levenberg-Marquardt algorithm \citep{More1979} to find many
local minima which each provide a good fit to the data and are
indisinguishable on statistical grounds.  All such statistically
indistinguishable local minima are shown in Figure \ref{fig:fits},
which was computed in the following manner. We first calculate the
observed (absorbed) flux with 90\% confidence error bars between 0.5
and 7 keV directly from the data using the CIAO \textit{srcflux}
command. At the first epoch the absorbed flux is $1.81^{0.9}_{-0.68}
\times 10^{-14}$ erg cm$^{-2}$ s$^{-1}$, and at the second epoch the
flux is $2.45^{+2.13}_{-1.35} \times 10^{-15}$ erg cm$^{-2}$
s$^{-1}$. { We then attempted to fit the ``vmekal'' model to the X-ray
  spectrum 7000 different times. Each fit had different initial
  guesses for $N_{\rm H}$ and $kT$. The initial guesses for $N_{\rm
    H}$ ranged from $0.0882 \times 10^{22}$ cm$^{-2}$ to $10 \times
  10^{22}$ cm$^{-2}$. The initial guesses for $kT$ ranged from 0.1 keV
  to 25 keV. We placed a lower bound on the column density ($N_{\rm
    H}$) of $0.0882 \times 10^{22}$ cm$^{-2}$, which is the Galactic
  column density in the direction of SN 2012ca
  \citep{DickeyLockman1990}. We also placed a conservative upper bound
  on the temperature of 25 keV, slightly higher than the proton
  temperature for a shock velocity of $3500$ km~s$^{-1}$.}

For each fit, the model-dependent absorbed flux was computed using
CIAO's \textit{calc\_energy\_flux} function. We rejected all fits for
which the model-dependent absorbed flux was not within the 90\%
confidence interval of the model-independent flux. We then visually
inspected the { fits} which passed this criteria to ensure that the {
  model} did not produce features which massively overestimated or
underestimated the data in any energy range, and rejected all fits
that gave a reduced test statistic exceeding 1. The final $N_{\rm H}$
and $kT$ values for acceptable fits are shown in Figure
\ref{fig:fits}, which we treated as the parameter space of suitable
values of $N_{\rm H}$ and $kT$.

From Figure \ref{fig:fits}, fits to the X-ray spectrum for the second
epoch cluster at low $kT$ and high $N_{\rm H}$. In contrast, the first
epoch appears to have values that occupy two distinct regions: a
region with low $N_{\rm H}$ and high $kT$, and a region with high
$N_{\rm H}$ and low $kT$. This provides a clue to understanding the
true best fits --- { since there is no obvious mechanism to add more
  material to the CSM as the SN evolves}, we would expect the second epoch to have a
lower value of $N_{\rm H}$ than the first, suggesting that the true
$N_{\rm H}$ at the first epoch must be high, combined with a low $kT$.

Another clue to the correct fitting parameters is provided by the
observed data. We can compute the electron number density $n_e$
directly from the X-ray luminosity $L = n_e^2\Lambda\,V$, given a
cooling function $\Lambda$ and the volume of the emitting region
$V$. { Here we compute the minimum possible density and infer a
  constraint on $N_{\rm H}$.} As a first guess, the minimum unabsorbed
luminosity $L$ is calculated from the unabsorbed flux, which we
determine from the observed count rate using the Galactic $N_{\rm H}$,
the lowest possible value of $N_{\rm H}$. This is done with the
Portable, Interactive Multi-Mission Simulator
(PIMMS\footnote{http://cxc.harvard.edu/toolkit/pimms.jsp}). { Because
  unabsorbed flux and thus density decreases with temperature}, a
temperature of 12 keV is used as a conservative value to minimize the
density calculation. The estimated unabsorbed fluxes are
$2.0^{+1.0}_{-0.7} \times 10^{-14}$ erg cm$^{-2}$ s$^{-1}$ for epoch 1
and $2.6^{+2.3}_{-1.5} \times 10^{-15}$ erg cm$^{-2}$ s$^{-1}$ for
epoch 2.  This procedure gives luminosities of $1.48^{+0.74}_{-0.55}
\times 10^{40}$ erg s$^{-1}$ for epoch 1, and $2.0^{+1.7}_{-1.1}
\times 10^{39}$ erg s$^{-1}$ for epoch 2. Note that $\Lambda$ is an
approximation of the cooling curve taken from \citet{cf94}.  The
highest observed optical velocity is 3200 km s$^{-1}$, taken here to
be the shock velocity, although it is possible that faster-moving
material exists that is not observed at optical wavelengths because it
has a lower emissivity, owing perhaps to a lower density. The radius
is calculated assuming a shock velocity of $3200 \pm 300$ km s$^{-1}$
and time since explosion. Since the inferred densities decrease with
increasing shock velocity, by assuming the maximum observed gas
velocity we minimize the inferred density. { The shock is assumed to
  be strong and nonradiative to begin with because this is the
  simplest scenario, which can be tested and if necessary modified for
  consistency once we obtain a value for the density}. We find that
the minimum $n_e$ is $\sim 5.8^{+4.73}_{-1.8} \times 10^6$ cm$^{-3}$
at epoch 1, depending on the exact temperature, width of the shocked
region, and shock velocity. The unshocked material has a density 4
times smaller for a strong shock. At epoch 2 the density of the
shocked material is $\sim 1.4^{+1.44}_{-0.58} \times 10^6$
cm$^{-3}$. The cooling time for these densities and this temperature
  is much longer than the length of time between the explosion and
  observation, so our assumption of a nonradiative shock is
  self-consistent. This represents the absolute minimum luminosity
and density.  The density gives us an idea of what the minimum value
of the column density should be, if integrated along the radius from
the first observation to the second. Clearly, this indicates a very
high column ${ > 5 \times 10^{21}}$ cm$^{-2}$.

Because we do not have spectra before 50 days after the explosion,
there is a possibility that the velocity of the shock was much higher
than assumed for these first 50 days. In the worst case, the velocity
is $10^{4}$ km s$^{-1}$, before sharply dropping to 3200 km
s$^{-1}$. This would mean that the volume of material in the density
calculation was underestimated, and thus our densities
overestimated. In this worst-case scenario, our inferred shocked
density for epoch 1 drops to $3.7^{+1.32}_{-1.05} \times 10^{6}$
cm$^{-3}$. For epoch 2, our inferred shocked density drops to
$9.4^{+5.23}_{-3.92} \times 10^{5}$ cm$^{-3}$. The column density must
still be $\ga 5 \times 10^{21}$ cm$^{-2}$.

\section{Calculating the Density of the CSM}

\subsection{Symmetric Medium} 
Keeping the above discussion in mind, we proceed to determine the
actual fit parameters using Figure \ref{fig:fits}. We used a kernel
density estimator with Gaussian kernels on the fits in Figure
\ref{fig:fits} to estimate the distributions for $N_{\rm H}$ and $kT$
for each epoch. We excluded all fits to epoch 1 in the high-$kT$ and
low-$N_{\rm H}$ regime, those with $kT > 6.4$ keV, because we have
already determined that $N_{\rm H} > 5 \times 10^{21}$ cm$^{-2}$. We
also excluded the fits to epoch 2 in the high-$kT$ and high-$N_{\rm
  H}$ region because $N_{\rm H}$ and $kT$ must be higher for epoch 1
than for epoch 2.  To optimize the width of the Gaussian kernel, we
performed the following procedure for a range of widths
\citep{ivezicetal14}. First, we removed one point from the dataset.
Next, we calculated the log of the likelihood of the density at the
removed point. We repeated this step for each point. The sum of each
log-likelihood divided by the number of fits was then used as a cost
function. This cost function was minimized with respect to the width
of the Gaussian kernel. We chose this cost function in order to
minimize the estimator's error between points, as the parameter space
for epoch 1 is sparsely populated. For epoch 1, the width of the $kT$
kernel was 1.8 keV and the width of the $N_{\rm H}$ kernel was ${
  1.6} \times 10^{22}$ cm$^{-2}$. For epoch 2, the width of the $kT$
kernel was 0.15 keV and the width of the $N_{\rm H}$ kernel was ${
  0.15} \times 10^{22}$ cm$^{-2}$. From our distributions of
temperature and column density, we calculated the expectation value
and 68\% confidence errors from the distribution. While the
uncertainties may appear small, we emphasize that the full range of
possible values is larger. A realistic value for epoch 1 is $kT =
2.63_{-1.82}^{+1.80}$ keV and $N_{\rm H} = 6.07_{-3.58}^{+3.46} \times
10^{22}$ cm$^{-2}$.

Taking the variation in all quantities into account, for epoch 1, this
gives a flux of $7.57_{-5.11}^{+389} \times 10^{-14}$ erg cm$^{-2}$
s$^{-1}$, and thus a luminosity of $ 5.81_{-3.92}^{+298} \times
10^{40}$ erg s$^{-1}$. We assume, despite not knowing the true shock
velocity, a strong nonradiative shock with velocity 3200 km
s$^{-1}$. We use the same formula as in \S \ref{sec:spectrum} to
compute the number density of the unshocked region, which we find to
be $3.46_{-1.44}^{+16.9} \times 10^6$ cm$^{-3}$ at epoch 1. The
cooling time for this temperature and density is $\sim 1340$ days,
greater than the length of time between the explosion and observation,
so the assumption of a nonradiative shock is self-consistent.

At epoch 2, we have $kT= 1.17_{-0.53}^{+0.37}$ keV and $N_{\rm H} =
5.20_{-2.20}^{+2.48} \times 10^{22}$ cm$^{-2}$, which gives an
unabsorbed flux of $3.68_{-3.05}^{+104.0} \times 10^{-14}$ erg
cm$^{-2}$ s$^{-1}$. This results in a luminosity of $2.82
^{+79.9}_{-2.34} \times 10^{40}$ erg s$^{-1}$ and an unshocked density
of $1.22^{+5.03}_{-0.74} \times 10^6$ cm$^{-3}$. The cooling time for
this temperature and density is $\sim 2340$ days, greater than the
length of time between the explosion and observation, so the
assumption of a nonradiative shock is self-consistent. Thus,
realistically, the density of the surrounding medium is $\sim 4 \times
10^6$ cm$^{-3}$ at epoch 1 and about a factor of 4 smaller at epoch 2.

Given an approximate mean molecular weight $\mu$ of 1, we can derive a
mass density $\sim 5.8^{+28.3}_{-2.4} \times 10^{-18}$ g cm$^{-3}$ for
epoch 1 and $\sim 2.0^{+8.5}_{-1.2} \times 10^{-18}$ g cm$^{-3}$ for
epoch 2. Taking the value of $n_e$ at the radius at epoch 1 ($r_1$),
and integrating $n_e$ along the line of sight from $r_1$ to the radius
at epoch 2 ($r_2$), assuming a constant density profile, we find that
$N_{\rm H} \approx 1.82 \times 10^{22}$ cm$^{-2}$. If we assume that
$n_e$ drops rapidly after $r_1$ to its value at $r_2$ and take the
value of $n_e$ at $r_2$ as more representative, then we get a value of
$N_{\rm H}$ lower by a factor of 3. In reality, it is probably
somewhere between the two, or possibly greater due to material beyond
$r_2$, and thus $ > 1 \times 10^{22}$ cm$^{-2}$. If we assume the
density at $r_2$ to be spread over the region all the way to at least
$r_2$, there is $\sim 0.04$ M$_{\odot}$ of material around the SN {
  between $r_1$ and $r_2$, a lower limit on the mass of the CSM}. The
density at $r_1$ increases this to 0.11 M$_{\odot}$. The actual value
is again probably somewhere in between, suggesting that at least $\sim
0.1 \pm 0.05$ M$_{\odot}$ of material is present around the SN.

In an ionised medium, the lower limit on $N_{\rm H}$ derived above may
not be valid if some or all of the material is ionised. It is
therefore worthwhile to check the ionisation state of the
medium. Since there is no nearby photoionising source, it is the X-ray
emission itself that tends to ionise the medium. We compute the
ionisation parameter given by \citet{KallmanMcCray1982}, $\xi =
{L_{X}}/{n_e r^2}$. An ionisation parameter $\xi \ga 100$ indicates that
elements such as C, N, and O are completely ionised, whereas
ionisation of heavier elements such as S and Fe requires $\xi \ga
1000$ \citep{ci12}. For epoch 1, $\xi = 71.43_{-38.15}^{+703.83}$. For
epoch 2, $\xi = 54.72_{-34.78}^{+325.20}$.  Our results show that the
gas is far from being fully ionised and therefore this should not
significantly affect the column density. Inferring the global
ionisation state from $\xi$ may be more difficult. For a spherically
symmetric medium and a steady-state wind (for example), $n_e \propto
r^{-2}$, and $\xi$ is a constant that describes the global state of
the medium. In the present scenario, we have no knowledge of the
actual density variation with radius, and the geometry of the medium
is likely not spherical as argued in the next section. In fact, using
the densities inferred for the high-density component in the next
section would reduce the ionisation parameter significantly,
indicating an almost neutral medium, with the caveat that the covering
fraction of the high-density component is unknown.

Since we are agnostic about the shock velocity for the first 50 days,
we must assume it is $10^{4}$ km s$^{-1}$ during that time and
calculate the density.  In this worst-case scenario, our inferred
density for epoch 1 drops to $2.16^{+10.12}_{-0.87} \times 10^{6}$
cm$^{-3}$. For epoch 2, our inferred density drops to $
8.44^{+33.79}_{-5.05} \times 10^{5}$ cm$^{-3}$. The inferred minimum
column density must still be $> 1 \times 10^{22}$ cm$^{-2}$.

\subsection{Asymmetry in the Medium}

The above calculations assume a spherically symmetric medium, which
would be the simplest assumption. The calculations do, however, lead to
a contradiction. The velocity indicated from the BSWZI, 3200 km
s$^{-1}$, implies a post-shock proton temperature of 12--20~keV,
depending on the mean molecular weight. The X-ray fits give an
electron temperature on the order of 2~keV or less. However, the high
densities implied by the above calculations would mean that Coulomb
equilibration should be important. For number densities greater than
$10^6$ cm$^{-3}$ and $T \approx 10^7$ K ($\sim 1$~keV), using Equation 3.6 
of \citet{Franssonetal1996}, the equilibration time is $< 10$ days. The
equilibration time for electrons and protons would be on the order of
2 days for epoch 1, so we would expect the electron and proton
temperatures to be equilibrated.

One way to resolve this is to assume that the X-ray emission is not
coming from the fastest-moving gas. This is not unreasonable, given
that there is evidence in the Gaussian line profiles for gas moving at
all velocities from $-3500$ km s$^{-1}$ to +3500 km s$^{-1}$. If most of
the emission was coming from a thin shell of gas with a small velocity
range, we would have seen more of a boxy line profile. This is best
exemplified by the case of SN 1993J \citep{franssonetal05}, where the
emission is inferred to arise from a thin shell having a small velocity
range. Instead, we see profiles that can be fit with a Gaussian
shape. Indeed, our high-resolution spectra suggest a FWHM of around
1300 km s$^{-1}$, indicating that there is sufficient material going
at several hundred to a thousand km s$^{-1}$.

This indicates the possible existence of a two-phase medium. Part of
the gas is traveling at speeds up to 3200 km s$^{-1}$, but is
expanding into a lower-density medium. Given the temperatures from the
X-ray fits, the X-ray emission arises from gas moving at velocities
around $1157^{+346}_{-518}$ km s$^{-1}$ during the first observation
and $773_{-219}^{+114}$ km s$^{-1}$ during the second
observation. Thus, the X-ray emission must be arising within a medium
where the shock velocity is lower than that of the forward shock, and
therefore the density is correspondingly higher.  In a
single-degenerate scenario, the dense medium may be due to clumps in
the surrounding environment, or a dense torus. A clumpy medium has
been inferred for SN 2002ic from spectroscopic \citep{Dengetal2004}
and spectropolarimetric \citep{wangetal04} observations.  Most
standard double-degenerate scenarios involving the merger of two white
dwarfs would not produce a dense medium, but one special case proposed
to explain H lines in SNe~Ia via a double-degenerate scenario
\citep{ss74, lr03} results in a circumbinary cloud or common envelope
around the SN. A surrounding disk is a possibility in both the single-
and double-degenerate scenarios \citep{ml90, hp06}, although it is
doubtful that the \citet{ml90} model could explain the H lines. A
circumstellar disk or torus is also an ingredient of binary evolution
models of SNe~Ia computed by \citet{hkn08}. If the clumps, torus,
envelope, or disk are in pressure equilibrium with the surrounding
material (which is not necessary in a transient situation), the
velocity ratios of a factor of 3--4 would indicate that this medium
was about 9--16 times denser than the interclump/interdisk medium.

The mass density of this high-density medium, $\rho_{\rm hd}$, can be
estimated. A $\sim 1000$ km s$^{-1}$ shock moving into a medium of
density $\sim 10^7$ cm$^{-3}$ would be radiative. Therefore, in this
scenario we use the equation for a radiative shock to estimate the
density, assuming that most of the emission from the radiative shock
arises at X-ray energies:

\be
L_X = 0.5 (\alpha \times 4 \pi r^2) \rho_{\rm hd} v_{\rm sh}^3, 
\ee

\noindent
where we take $v_{sh}$ to be the velocity associated with the
temperature given by the X-ray fits, and $\alpha \times 4 \pi r^2$ is
the area of the emitting region, taken as a fraction $\alpha \le 1$ of
the spherical emitting area $4 \pi r^2$ . For epoch 1, $v_{\rm
  sh}=1157^{+346}_{-518}$ km s$^{-1}$, and for epoch 2, $v_{\rm
  sh}=773_{-219}^{+114}$ km s$^{-1}$. The radius, or the surface area,
of the clumps/disk is not known. Given the X-ray temperatures we
obtain, we set the radius assuming a constant velocity of $
1157^{+346}_{-518}$ km s$^{-1}$ over 554 days. For the radius of the
second observation, we assume the same constant velocity for the first
554 days and then a velocity of $773_{-219}^{+114}$ km s$^{-1}$ out to
745 days.  We find $\rho_{\rm hd}= 1.95^{+1977}_{-1.78} \times
10^{-16} {\alpha}^{-1}$ g cm$^{-3}$ and $\rho_{\rm
  hd}=3.10^{+562}_{-2.86} \times 10^{-16} {\alpha}^{-1}$ g cm$^{-3}$
for the first and second observations, respectively. These density
values correspond to number densities of $1.16^{+1181}_{-1.06} \times
10^8 {\alpha}^{-1}$ cm$^{-3}$ and $1.85^{+336}_{-1.70} \times 10^{8}
{\alpha}^{-1}$ cm$^{-3}$ for the first and second observations,
respectively. If the area is larger, the density will decrease
correspondingly, and vice versa.  The cooling time of both a 1157 km
s$^{-1}$ shock and a 773 km s$^{-1}$ shock into a region with such a
density is less than twenty days, so our assumption of a radiative
shock is justified.

This model fits some aspects of the data better, and so may be
preferable. While the lack of detailed information does not allow for
accurate calculations, it is clear that the densities are possibly
even higher than in the symmetrical case, as expected. However, they
are consistent with densities suggested for other SNe~Ia-CSM
\citep{wangetal04, Dengetal2004,Alderingetal2006}.  These high
densities are also consistent with those required for collisional
excitation to be responsible for the high Balmer decrement values
\citep{du80}, which require $n_e > 10^8$ cm$^{-3}$.

\section{Discussion and Conclusions}
We describe here the X-ray emission from SN 2012ca, the first Type
Ia-CSM SN to be detected in X-rays. Although the statistics preclude
an accurate fit, there are several indicators that the SN is expanding
in an extremely high density medium, with density exceeding $10^6$
cm$^{-3}$, and $> 10^8$ cm$^{-3}$ in our preferred scenario. The high
Balmer decrement seen in this SN, with values ranging from 3 to 20,
has been interpreted in other SNe~Ia-CSM as being most likely produced
through excitation rather than recombination
\citep{silvermanetal2013a}, which requires a very high density medium,
on the order of $n_e \approx 10^8$ cm$^{-3}$. The spectral fits to the X-ray data imply
a large column density $> 10^{22}$ cm$^{-2}$, consistent with a large
CSM density. The relatively low shock velocity (3200 km s$^{-1}$ at
early times) also suggests a density orders of magnitude higher than
that of a normal SN~Ia.

Although the data could be fit by a spherically symmetric medium, this
does lead to a contradiction in that the electron and ion temperatures
are quite different, whereas the low Coulomb equilibration time
suggests that they should be similar.  The difference between the
post-shock proton temperature and the fitted electron temperatures
could indicate that the X-ray emission does not arise from the
fast-moving gas, but from slower-velocity gas expanding into a denser
medium. The fact that the optical spectra indicate there is a large
amount of gas moving at low velocities suggests that our low X-ray
temperatures are correct. This dense medium could consist of
high-density clumps, a dense torus, a dense equatorial disk, or a
common envelope.  The X-ray emission would arise from shocks, probably
radiative, entering this dense medium. The existence of this dense
medium renders most double-degenerate scenarios unviable, except
perhaps for one which involves the degenerate core of an asymptotic
giant branch (AGB) star that
shed its H envelope in a merger with a companion white dwarf shortly
before the explosion \citep{lr03}.

The BSWZI of the H$\alpha$ emission indicates that it arises from the
main 3200 km s$^{-1}$ shock. This, however, presents a problem: it is
unlikely that a nonradiative shock could give rise to such a level of
Balmer emission. Using \citet{cr78}, the ratio of H$\alpha$ power to
shock power is of order 10$^{-6}$, which is much smaller than the
observed H$\alpha$ luminosity \citep{Foxetal2015}. The emission is
probably not due to a spherical shock--- as noted earlier, this would
result in a boxy line profile at the shock velocity that is not
seen. A radiative shock could be inferred, but if we assume the X-ray
luminosity to be the shock luminosity, the density inferred due to the
high shock velocity (from Equation 1) is not enough to cool the shock
in the required time.  Thus, none of these solutions is particularly
attractive, and there is an inherent lack of self-consistency in some
of them. This suggests that the interaction is probably more complex
than noted here, occurring presumably at different velocities and with
variable-density ambient medium. We note that a similar problem also
occurs in many SNe~IIn.

\citet{Inserraetal2016} use the optical light curves to compute the
mass of the CSM in SN 2012ca, finding 2.3--2.6 M$_\odot$.  For the
spherically symmetric case, our mass is only 0.1 M$_\odot$, much lower
than their value. However, in the case of an asymmetric or clumpy
medium, the densities of the high-density component are about two
orders of magnitude greater. If this higher-density component has even
a 10\% filling factor, then we would expect a CSM mass in the
asymmetric-medium model of $\sim 1$ M$_\odot$, which is more
consistent with the value of \citet{Inserraetal2016}. Thus, an
additional factor in favor of the asymmetric-medium model is the
higher CSM mass, which is more likely able to power the optical
luminosity. The discrepancy in the computed masses is not significant,
considering that \citet{Inserraetal2016} use an ejecta profile that is
not appropriate for Type Ia SNe \citep{dc98}, and they do not make
allowances for deviations from spherical symmetry and a clumpy
medium. Their model also requires a kinetic energy input of 7--9
$\times 10^{51}$ erg, which is large even compared to that of other
SNe~Ia-CSM modelled by them.

While we are constrained by the available data, in \S
\ref{sec:spectrum} we computed the absolute minimum flux and thereby
density, assuming merely the observed flux and the Galactic column
density. We reiterate that these provide a minimum limit to the
ambient number density of $> 10^6$ cm$^{-3}$, and show clearly that it
is still higher than is typical for most SNe after 1.5~yr, and in the
range deduced for SNe~Ia-CSM in general. In order for all the
observations to be consistent thereafter, the final deduced density
ranges seem quite appropriate.

\begin{figure}
\center
\includegraphics[width=0.4\textwidth, angle=-270]{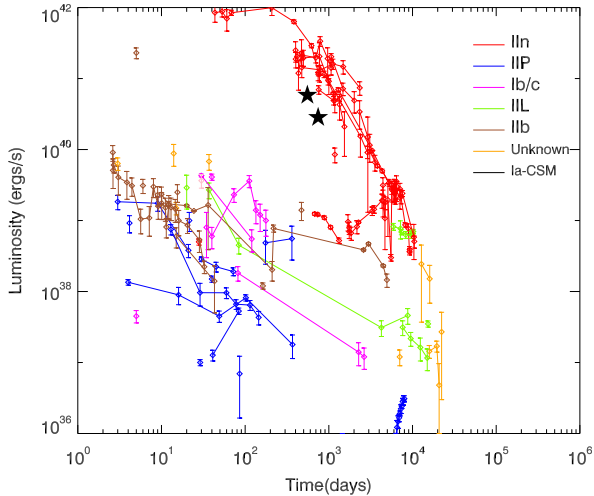}
\caption{The X-ray light curves of most X-ray SNe, grouped by
  type. Adapted from \citet{vvd14}, with SN 2012ca and some other data
  points added. The stars represent the nominal X-ray luminosities at
  the two epochs for SN 2012ca. Note that these place it in the middle
  of the range for SNe~IIn.
  \label{fig:lc}}
\end{figure}

The inferred densities from SN 2012ca are extremely high compared to
those of other SNe~Ia, which typically expand in much lower
densities. However, they are consistent with those inferred for other
members of the SN~Ia-CSM subclass, with most showing indications of
high-density surroundings
\citep{silvermanetal2013a}. \citet{Alderingetal2006} suggest that SN
2005gj, also identified as a SN~Ia-CSM, had an ambient density $n_e >
10^{8}$ cm$^{-3}$.  \citet{wangetal04} suggest that SN 2002ic had a
dense, clumpy, disk-like environment, with clumps of density $> 10^8$
cm$^{-3}$ and sizes $5 \times 10^{16}$ cm. This is quite similar to
the picture envisioned here for SN 2012ca. \citet{Dengetal2004} also
suggest a dense, clumpy, and aspherical circumstellar medium for SN
2002ic, with mass-loss rates as high as $10^{-2}$ v$_{w100}$
M$_{\odot}$ yr$^{-1}$, where v$_{w100}$ = v$_w/100 \;{\rm km
  \;s^{-1}}$, and v$_w$ is the stellar wind velocity. The density is
much higher than that around typical core-collapse SNe, and perhaps
also the subclass of Type IIn SNe. Densities of up to $n_e \approx
10^{6}$ cm$^{-3}$ have been estimated for the Type IIn SN 2006jd
\citep{ccc12}, and between $3 \times 10^6$ and $10^8$ cm$^{-3}$ for SN
2010jl \citep{franssonetal14}. High mass-loss rates have been found
for SN 2005kd \citep{drrb16} and SN 2005ip \citep{Katsudaetal2014}. A
clumpy medium has been suggested for Type IIn SNe such as SN 1988Z, SN
1978K, and SN 1986J \citep{chugai93,cd94,cdd95}, and a dense torus for
SNe 2005kd, 2006jd, and 2010jl \citep{katsudaetal16}. SNe~IIn in
general have high X-ray luminosities \citep[Figure
  \ref{fig:lc};][]{vvd12, vvd14, chandraetal15, drrb16}, which if due
to thermal emission suggest high densities. Note that the luminosity
predicted for SN 2012ca, of order $10^{40}$ erg s$^{-1}$, places it
squarely within the range of the X-ray luminosities of SNe~IIn
\citep[Figure \ref{fig:lc};][]{vvd12,drrb16}. { Despite these
  similarities, the temperature of SN 2012ca is significantly lower
  than that of most SNe~IIn at a similar epoch. It typically takes SNe~IIn a
  few thousand days to reach such low temperatures. The high CSM
  density around SN 2012ca may allow it to cool more quickly,
  explaining this discrepancy.} Thus, while the spectra suggest a Type
Ia SN, the X-ray luminosity, high density, and circumstellar
interaction are typical of a core-collapse Type IIn SNe, suggesting a
CSM similar to that seen in SNe~IIn.

While SN 2012ca is the first Type Ia-CSM SN to be detected in X-rays,
others have been examined unsuccessfully for signs of X-ray
emission. \citet{hughesetal07} took deep X-ray observations of SN
2002ic and SN 2005gj and placed upper limits on their X-ray flux.  The
limits on SN 2005gj are more stringent. The redshift 0.019 of SN
2012ca places it much closer than SN 2002ic ($z = 0.0667$) and SN
2005gj ($z = 0.062$). If SN 2012ca had occurred at the redshift of SN
2002ic, it would not have been detected.  If it had occurred at the
redshift of SN 2005gj, it may have been detected if it was cooler than
10 keV and the absorption of X-ray flux by the CSM was not
significant. However, given the density estimate by
\citet{Alderingetal2006}, absorption in SN 2005gj could be higher than
in SN 2012ca, making it undetectable. Furthermore, these SNe were
observed at earlier epochs than SN 2012ca: SN 2002ic was observed 275
days after the explosion, while SN 2005gj was observed merely 81 days
after the explosion. The column densities would have been much larger
for similar external densities, thus contributing to making them
undetectable.

It is not clear how the surrounding medium is produced, and
furthermore what its density profile looks like. White dwarfs are not
expected to experience any mass loss or have a high-density medium
surrounding them. However, there are several scenarios for
single-degenerate progenitors which produce an asymmetric dense medium
via the companion. Single-degenerate scenarios could produce a clumpy
medium, a dense torus, or a dense disk. Double-degenerate scenarios
would not generally be expected to produce a H-rich dense medium,
although one model does predict a H-rich circumbinary cloud or common
envelope.

If the density profile is due to a stellar wind with constant
mass-loss parameters $\rho_{\rm CSM}={\dot{M}}/{4\pi r^{2} v_w}$,
where $\dot{M}$ is the mass-loss rate and $v_w$ the wind velocity,
then for epoch 1, $\dot{M}/v_w$ must be larger than $2.7 \times
10^{-4}$ v$_{w10}$ M$_{\odot}$ yr$^{-1}$ in the spherically symmetric
case, and two orders of magnitude larger in the two-component medium
case (although asphericity may increase it further). Despite being
extremely high, this mass-loss rate is consistent with those derived
from light-curve modelling of SNe 2002ic and 1997cy \citep{cy04}, as
well as that derived from spectral modelling of SN 2002ic
\citep{Kotaketal2004}. This lower limit is orders of magnitude larger
than upper limits derived from deep X-ray observations of nearby
SNe~Ia. Margutti et al.~(2012) place an upper limit on $\dot{M}/v_w$
for the nearby Type Ia SN 2011fe of $2 \times 10^{-10}$ v$_{w10}$
M$_{\odot}$ yr$^{-1}$, while Margutti et al.~(2014) posit an upper
limit on SN 2014J of $2.5 \times 10^{-10}$ v$_{w10}$ M$_{\odot}$
yr$^{-1}$. Of course, the constant velocity over at least 500 days
suggests that if there is a wind medium, its density is not decreasing
as steeply as $r^{-2}$ but more gently \citep{vvd12}.

This mass-loss rate may only be satisfied by the higher end of red
supergiant stars \citep{mj11} and by yellow hypergiant stars. If the
surrounding velocity is higher, say 100-1000 km s$^{-1}$, then the
mass-loss rate increases accordingly. This would then require a
luminous blue variable (LBV) undergoing eruptive mass loss
\citep{smith14}. Other SN progenitors such as Wolf-Rayet (W-R) stars
have winds of $\sim 2000$ km s$^{-1}$, and would therefore require a
mass-loss rate 100 times higher than that postulated above, which is
orders of magnitude higher than those actually measured for W-R stars
\citep{moffat15}. In addition, the presence of an H-rich medium
immediately around the star effectively rules out W-R stars.

It must be pointed out here that having a high-mass secondary likely
seems implausible around a white dwarf, given that one would have
expected the higher-mass companion to have evolved faster and
therefore be long gone before the white dwarf made an appearance. The
only lower-mass ($< 8$ M$_{\odot}$) progenitor that may be able to
satisfy the high mass-loss rates for at least a short period would be
an asymptotic giant branch star. 

The ambient medium may not result from a freely expanding wind at all,
but could be the result of a swept-up medium due to interacting winds
\citep{vvd11}, as suggested for SN 1996cr \citep{ddb10}. In this case,
the two winds do not need to have very high mass-loss rates, as long
as there is sufficient mass for the later wind to sweep up. However,
this would mean a low density very near the star, yet the low velocity
shortly after explosion seems to argue against this.

An ambient medium owing to recurrent nova eruptions was suggested by
\citet{Dildayetal2012} for PTF11kx. In that case, the presence of
multiple components, with fast-moving material inside denser and
slower-moving shells, could be explained by the SN shock interacting
with shells of material resulting from episodic nova eruptions. It is
not clear if such a model could apply here, since the SN shock
maintains a low velocity that does not vary much over the first 550
days (see Figure 1).

Kepler's SN remnant is a possible Galactic example of a SN~Ia that
requires dense mass loss. However, the CSM of Kepler's SN remnant is
much farther out than that of SN 2012ca \citep{Reynoldsetal2007,
  katsudaetal15}. \citet{katsudaetal15} argue that Kepler's SN was a
highly overluminous SN~Ia, and that the CSM consists of tenuous gas
with dense knots, similar to the model outlined herein. Their analysis
suggests that the knots were parsecs away from the progenitor, and the
medium shows evidence of CNO processing --- so, although both require
a SN~Ia progenitor and dense surrounding medium, the two events are
not directly comparable. They infer a high mass-loss rate of $10^{-5}$
yr$^{-1}$, which is considerably larger than the upper limits quoted
above, although lower than that inferred for SN 2012ca, especially in
the asymmetric-medium scenario. Their models require a progenitor with
a high mass-loss rate and no surviving companion; a recurrent-nova
scenario is not favored for Kepler's SN. The core-degenerate scenario
laid out by \citet{lr03} and elaborated further by
\citet{TsebrenkoSoker2013} could be the origin of Kepler's SN
remnant. The supersoft channel investigated by \citet{hp06} also
remains a possibility.

Finally, as noted earlier, it is possible that this SN, and others
like it, are not part of the class of SNe~Ia at all. They most
closely resemble SNe~IIn, which presumably have more than one
progenitor, all of the core-collapse variety.  That, however, then
leads to the question of why spectra of SNe~Ia-CSM resemble those of 
the SN~Ia class the most, and how the explosion of a massive star could 
give rise to SN~Ia-like spectral features.

Notwithstanding its origin, the high density surrounding SN 2012ca is
indisputable. The SN is fading in X-rays, and unlikely to be
detectable any more by current instruments. However, it makes a clear
case for other SNe~Ia-CSM to be observed in X-rays, and presumably at
radio wavelengths as well, opening up a new window to understanding
this class of intriguing objects and unraveling the mystery of their
progenitors. { Continued long-term observations would also be
  useful to show if SNe of this class eventually evolve into
  remnants resembling that of Kepler's SN}.

\section*{Acknowledgments}
V.V.D.'s research is supported by NASA Astrophysics Data Analysis
program grant NNX14AR63G (PI Dwarkadas) awarded to the University of
Chicago. J.M.S. is supported by an NSF Astronomy and Astrophysics
Postdoctoral Fellowship under award AST-1302771. O.D.F. was partially
supported by {\it Chandra} grant GO4-15052X provided by NASA through
the {\it Chandra} X-ray Observatory center, operated by SAO under NASA
contract NAS8-03060. A.V.F. has been supported by the Christopher
R. Redlich Fund, the TABASGO Foundation, NSF grant AST-1211916, and
the Miller Institute for Basic Research in Science (UC Berkeley).  His
work was conducted in part at the Aspen Center for Physics, which is
supported by NSF grant PHY-1607611; he thanks the Center for its
hospitality during the neutron stars workshop in June and July
2017. This research has made use of data obtained from the {\it
  Chandra} Data Archive, and software provided by the {\it Chandra}
X-ray Center (CXC) in the application packages CIAO, CHIPS, and
SHERPA. { We would like the thank the anonymous referee for a helpful
  and thorough reading of this paper.}

\bibliographystyle{mn2e}

\bibliography{bibliography_v19}

\bsp

\label{lastpage}

\end{document}